# Stellar Population Astrophysics (SPA) with TNG*

## Fluorine abundances in seven open clusters

S. Bijavara Seshashayana,[1] H. Jönsson,[1] V. D'Orazi,[2,3] G. Nandakumar,[4] E. Oliva,[5] A. Bragaglia,[6] N. Sanna,[5] D. Romano,[6] E. Spitoni,[7] A. Karakas,[8] M. Lugaro,[9,8] L. Origlia[6]

[1] Materials Science and Applied Mathematics, Malmö University, SE-205 06 Malmö, Sweden
   e-mail: `shilpa.bijavara-seshashayana@mau.se`
[2] Department of Physics, University of Rome Tor Vergata, via della Ricerca Scientifica 1, 00133 Rome, Italy
[3] INAF-Osservatorio Astronomico di Padova, vicolo dell' Osservatorio 5, 35122 Padova, Italy
[4] Lund Observatory, Division of Astrophysics, Department of Physics, Lund University, 22100 Lund, Sweden
[5] INAF-Osservatorio Astrofisico di Arcetri, Largo Enrico Fermi 5, 50125 Florence, Italy
[6] INAF-Osservatorio di Astrofisica e Scienza dello Spazio di Bologna, via Piero Gobetti 93/3, 40129 Bologna, Italy
[7] INAF-Osservatorio Astronomico di Trieste, via G.B.Tiepolo 11, 34131 Trieste, Italy
[8] School of Physics & Astronomy, Monash University, Clayton VIC 3800, Australia
[9] Konkoly Observatory, Research Centre for Astronomy and Earth Sciences (CSFK), ELKH, Konkoly Thege M. ´ut 15–17., H-1121 Budapest, Hungary

Received ; accepted

## ABSTRACT

*Context.* The age, evolution, and chemical properties of the Galactic disk can be effectively ascertained using open clusters. Within the large program Stellar Populations Astrophysics at the Telescopio Nazionale *Galileo*, we specifically focused on stars in open clusters, to investigate various astrophysical topics, from the chemical content of very young systems to the abundance patterns of lesser studied intermediate-age and old open clusters.
*Aims.* We investigate the astrophysically interesting element fluorine (F), which has an uncertain and intriguing cosmic origin. We also determine the abundance of cerium (Ce), as F abundance is expected to correlate with the s-process elements. We intend to determine the trend of F abundance across the Galactic disk as a function of metallicity and age. This will offer insights into Galactic chemical evolution models, potentially enhancing our comprehension of this element's cosmic origin.
*Methods.* High-resolution near-infrared spectra were obtained using the GIANO-B spectrograph. The Python version of Spectroscopy Made Easy (PySME), was used to derive atmospheric parameters and abundances. The stellar parameters were determined using OH, CN, and CO molecular lines along with Fe I lines. The F and Ce abundances were inferred using two K-band HF lines ($\lambda\lambda$ 2.28, 2.33 $\mu$m) and two atomic H-band lines ($\lambda\lambda$ 1.66, and 1.71 $\mu$m), respectively.
*Results.* Of all the clusters in our sample, only King 11 had not been previously studied through medium- to high-resolution spectroscopy, and our stellar parameter and metallicity findings align well with those documented in the literature. We have successfully inferred F and Ce abundances in all seven open clusters and probed the radial and age distributions of abundance ratios. This paper presents the first F Galactic radial abundance gradient. Our results are also compared with literature estimates and with Galactic chemical evolution models that have been generated using different F production channels.
*Conclusions.* Our results indicate a constant, solar pattern in the [F/Fe] ratios across clusters of different ages, supporting the latest findings that fluorine levels do not exhibit any secondary behavior for stars with solar or above-solar metallicity. However, an exception to this trend is seen in NGC 6791, a metal-rich, ancient cluster whose chemical composition is distinct due to its enhanced fluorine abundance. This anomaly strengthens the hypothesis that NGC 6791 originated in the inner regions of the Galaxy before migrating to its present position. By comparing our sample stars with the predictions of Galactic chemical evolution models, we came to the conclusion that both asymptotic giant branch stars and massive stars, including a fraction of fast rotators that increase with decreasing metallicity, are needed to explain the cosmic origin of F.

## 1. Introduction

In the field of Galactic archaeology, fluorine (F) stands out as a particularly captivating element due to its mysteri-ous origins. It accounts for less than one percent of the solar abundance much less than, for example, its periodic table neighbors carbon (C), oxygen (O), and nitrogen (N) making it only the $24^{th}$ most abundant element in the Universe. Gaining insights into the origins and production of F can help us significantly refine our models of chemical and stellar evolution. Interestingly, the production mechanisms for F may bear similarities to those of heavier elements such as Sr, La, and Ce in asymptotic giant branch (AGB) stars (see Ryde et al. 2020 and references therein) and also massive stars (Frischknecht et al. 2012; Prantzos et al. 2018). Observational challenges in measuring the cosmic content of F are due to its limited abundance, which results in weak spectral signatures. Additionally, strong atomic lines are sparse at both infrared and optical wavelengths, with only a few detectable infrared rotational hydrogen fluoride (HF) molecular lines, which are often contaminated by telluric lines.

---

* Based on observations made with the Italian Telescopio Nazionale Galileo (TNG) operated on the island of La Palma by the Fundación Galileo Galilei of the INAF (Istituto Nazionale di Astrofisica) at the Observatorio del Roque de los Muchachos. This study is part of the Large Program titled SPA – Stellar Population Astrophysics: a detailed, age-resolved chemical study of the Milky Way disk (PI: L. Origlia), granted observing time with HARPS-N and GIANO-B echelle spectrographs at the TNG.

That being said, the cosmic origin of F has been narrowed down to one or more of five sources:

- High-mass stars: In the case of massive stars, F can be produced in various ways. Rapidly rotating massive stars can be the source of primary F from $^{14}$N (Prantzos et al. 2018). In the presence of $^{13}$C, primary F is created from $^{14}$N via proton and $\alpha$ capture. $^{14}$N is the result of reactions with $^{12}$C, which is created from the burning of He in the massive star and is, therefore, primary (Guerço et al. 2019). F can also be produced by the massive stars that evolve as rare Wolf-Rayet (W-R) stars and may contribute significantly to the cosmic budget of F (Meynet & Arnould 1993). They are the source of powerful metallicity-dependent, radiatively driven stellar winds that can act as a shield against F destruction. As the convective core shrinks, the destruction of F via the ($\alpha$, p) reaction is reduced (Meynet & Arnould 1993, 2000). However, Palacios et al. (2005) suggested that the proposed production of F in W-R stars might be erroneous since when more recent yields are integrated and models of rotating W-R stars are included, the production of F from these stars is found to be significantly lower. And finally, F can be produced by the $\nu$ process in Type II supernovae (SNe II): the massive neutrino flux produced when an SNe II explodes (Woosley et al. 1990; Kobayashi et al. 2011b), may, despite the small cross-section (Woosley & Haxton 1988) react with $^{20}$Ne to produce F. The estimated $\nu$ energy is $E_\nu = 3 \times 10^{53}$ erg (Hartmann et al. 1991), but the exact value of this energy turns out to be important for the production of F, as studied by Kobayashi et al. (2011a). They conclude that the relative contribution of the $\nu$-process is greatest at low metallicities.

- Low-mass stars: In AGB stars, thermal pulses trigger the production of both F and the s-process elements. $^{14}$N is used as the "seed nucleus" to produce F via $\alpha$, neutron, and proton captures, and $^{13}C(\alpha,n)^{16}O$ is the primary neutron source (Busso et al. 1999). The models suggest that F is formed in AGB stars through the chain of reactions $^{14}N(n,p)^{14}C(\alpha,\gamma)^{18}O(p,\alpha)^{15}N(\alpha,\gamma)^{19}F$ (Lugaro et al. 2004; Cristallo et al. 2014), and, subsequently, is brought to the envelope via the third dredge up (Mowlavi et al. 1998; Goriely & Mowlavi 2000; Abia et al. 2009). There have been observations of the production of F in AGB stars (see, e.g., Jorissen et al. 1992; Lucatello et al. 2011; Abia et al. 2015, 2019). Finally, in novae, the mechanism that produces F is $^{17}O(p,\gamma)^{18}F(p,\gamma)^{19}Ne(\beta^+)F$. However, the production of F in novae and the $\nu$ process are largely uncertain (Kobayashi et al. 2011a).

The origin of F continues to be a subject of debate, with various theories persisting despite numerous attempts to unravel its mysteries using diverse methods and models. Renda et al. (2004) find that AGB stars dominate the F production in the early ages of the Galaxy's evolution due to the metallicity-dependent behavior of the AGB models, and W-R stars are the major contributors at solar and supersolar metallicities due to the metallicity-dependent mass-loss prescription used in the stellar models. However, this contradicts Olive & Vangioni (2019), who propose that AGB stars dominate F production at high metallicities, and the $\nu$ process dominates at low metallicities. The effects of metallicity, mass loss, and rotation were presented for the first time by Prantzos et al. (2018). Galactic chemical evolution (GCE) models that incorporate these findings were later used in Grisoni et al. (2020) to study the evolution of F in both thin- and thick-disk components. They concluded that rotating massive stars are the main contributors to F evolution at solar metallicities. Spitoni et al. (2018) mention the importance of including novae, which helps in reproducing the secondary behavior of F by a large margin. In Womack et al. (2023), novae and W-R stars were excluded as one of the prominent contributors in the chemical evolution of F. The observed patterns of [F/O] versus [O/H] were reproduced without a major contribution from novae. The wind yield was six times smaller than the core collapse ejecta in W-R stars. Despite the multitude of potential production sites, assessing F content continues to pose difficulties.

To clarify this intricate scenario, we aim to chart the F chemical evolution of the Galactic disk through open clusters (OCs). Over recent years, these stellar systems have been pivotal in our exploration of the formation and evolution mechanisms of stars and galaxies (e.g., Viana Almeida et al. 2009; D'Orazi et al. 2011; Yong et al. 2012; Magrini et al. 2017, 2023; Casamiquela et al. 2019; Spina et al. 2022; Myers et al. 2022). OCs are ideal tracers because they cover a significant range of galactocentric distances ($R_{gc}$), metallicities, and ages (see Spina et al. 2022 for a recent review). Furthermore, the age determination of OCs is much less uncertain than that of field stars, since isochrone-fitting methods can be applied to the entire ensemble of stars simultaneously instead of on a star-by-star basis, as is the case for field stars. Extensive spectroscopic campaigns, such as e.g., *Gaia*−European Southern Observatory (*Gaia*−ESO), GALactic Archaeology with HERMES (GALAH), and the Apache Point Observatory Galactic Evolution Experiment (APOGEE) (Gilmore et al. 2022; Randich et al. 2022; De Silva et al. 2015; Majewski et al. 2017) encompass a considerable number of OCs. These surveys are reshaping our comprehension of individual star evolution, binary systems, nucleosynthesis, and the formation and chemical evolution of the Galactic disk(s). Nonetheless, because they lack suitable wavelength coverage (and spectral resolution), none of these current (or upcoming, e.g., 4MOST, WEAVE, and MOONS; de Jong 2019; Jin et al. 2023; Gonzalez et al. 2020) stellar surveys will be able to determine F abundances. In this paper we unveil the first findings of our project, which aims to ascertain F abundances in OCs by utilizing the high-resolution near-infrared (NIR) spectrograph *GIANO-B*. This dataset forms a segment of the Stellar Population Astrophysics (SPA) project (PI L. Origlia) conducted at the Italian Telescopio Nazionale Galileo (TNG). Concurrently, specific new observations were acquired under the project FLUO (FLUorine abundances in Open cluster cool giants), which started in August 2023 and will be presented in a forthcoming paper. The present analysis encompasses a sample of seven OCs: NGC 7044, Ruprecht 171, Trumpler 5, King 11, NGC 7789, NGC 6819, and NGC 6791. We delve into the F and Ce abundance correlations with the cluster $R_{gc}$ and age. Our project will provide the largest homogeneous database for the determination of F in OCs.

## 2. Observations

We acquired all our data at the 3.6 m, Italian TNG located at the Observatorio del Roque de los Muchachos on La Palma (Canary Islands, Spain). The two high-resolution spectrographs, High Accuracy Radial Velocity Planet Searcher for the Northern hemisphere (HARPS−N) (Optical, $R = 115000$, $\lambda\lambda = 3800 - 6900$Å, Cosentino 2014) and GIANO-B (NIR; $R = 50000$, $\lambda\lambda = 0.97 - 2.5\mu m$), were used in GIARPS (GIANO-B + HARPS-N) mode (Oliva et al. 2012a,b; Origlia 2014). The optical and NIR observations were carried out simultaneously by separating the light using a dichroic. In this paper, however, only the NIR data are exploited. The observations were carried out between August 2018 and December 2021. GIANO-B was used to collect the spectra by nodding the star along the slit in which the targets are positioned at 1/4 (position A) and 3/4 (position B) of the slit length. The spectra were reduced using the software GOFIO (Rainer 2018). This includes bad pixel removal, sky and dark subtraction, flat-fielding, optimal spectrum extraction, and wavelength calibration. Telluric contamination was removed using a standard, early-type star observed during the same night, following the standard strategy described in for example, Ryde et al. (2019).

Our sample comprises 17 stars in seven OCs, namely NGC 7044, Ruprecht 171, Trumpler 5, King 11, NGC 7789, NGC 6819, and NGC 6791 (see Table 1).

## 3. Analysis

We used the Python version of Spectroscopy Made Easy (PySME) to analyze our stellar spectra (Piskunov & Valenti 2017; Valenti & Piskunov 1996; Wehrhahn et al. 2023). It uses $\chi^2$ minimization to fit calculated synthetic spectra to the observed spectra in pre-selected regions, using all the information provided along with the desired parameters. MARCS model atmospheres were used (Gustafsson et al. 2008) along with line lists taken from the Vienna Atomic Line Database (VALD3) (Piskunov et al. 1995; Kupka et al. 1999, 2000; Ryabchikova et al. 2015). Non-local thermodynamic equilibrium (NLTE) corrections were used for atomic spectral lines for several elements e.g., Na, Mg, Al (see Amarsi et al. 2020, for details), but the only lines relevant for this paper that were synthesized in NLTE are Fe I lines.

The useful lines of the HF molecule are in the K and N bands at $2.1 - 2.4\mu m$ and $8 - 13\mu m$, respectively; only the first is available in the GIANO-B spectra. These lines have been observed in cool giants with $T_{eff} < 4500$ K (e.g., Jorissen et al. 1992), although telluric lines can heavily affect the abundance determination (de Laverny & Recio-Blanco 2013). Nonetheless, the HF line list is now well established (Jönsson et al. 2014a,b). The lines used to determine both F and Ce abundances are given in Table 2. For F, the molecular lines at 22778.249Å and 23358.329Å were taken from Jönsson et al. (2014a). Nandakumar et al. (2023b) demonstrated that different lines of the HF spectrum could result in different inferred abundances for the same stars, with significant trends as a function of effective temperature and (high) metallicity. In particular, they showed that when using the most commonly adopted line at 23358.329Å, there is a significant upward trend for cool ($T_{eff} < 3500$ K) and metal-rich ([Fe/H] > 0.0 dex) stars (Nandakumar et al. 2023b). Therefore, the line at 23358.329Å was omitted in the cool, metal-rich cluster NGC6791 in our analysis. Atomic parameters for Ce II lines at 16595.180Å and 17058.880Å were retrieved from Corliss (1973). The $\log(gf)$ values given for Ce were originally determined astrophysically by Cunha et al. (2017), but have been slightly adjusted so that the averaged abundances derived from them agree with the measurements from optical spectra of the same solar neighborhood stars (Montelius et al. 2022); see our Table 2.

As a representative example, the GIANO-B spectra covering the HF and Ce lines for two stars in each of NGC7044 and NGC6819 are shown in Fig. 1. The solid red line is the synthetic spectra and black dots are for the observed spectra.

### 3.1. Stellar parameters

The fundamental stellar parameters, namely the effective temperature ($T_{eff}$), surface gravity ($\log g$), metallicity ([Fe/H]), and microturbulence velocity ($v_{mic}$), are crucial for deriving elemental abundances using spectrum synthesis methods. In this work, we used an iterative method to determine the fundamental stellar parameters using a selected set of $T_{eff}$ sensitive molecular OH lines, along with Fe atomic lines, CN, and CO molecular band heads in the H band wavelength regime (14000 - 18000 Å). We give a brief overview of the method below, the details of which can be found in Nandakumar et al. (2023a).

In this method, firstly, $T_{eff}$, [Fe/H], $v_{mic}$, macroturbulence velocity ($v_{mac}$), C and N were determined from the spectra assuming a $\log g$ and an oxygen abundance. Then, the $\log g$ and oxygen abundances were updated, and the same spectral analysis was redone until the change in values between iterations was negligible. We assumed oxygen abundances based on the [O/Fe] versus [Fe/H] trends for the disk of the Milky Way from Amarsi et al. (2019) (left panel of their Fig. 12), and $\log g$ for a combination of $T_{eff}$ and [Fe/H] is determined from the Yonsei-Yale (YY) isochrones (Demarque et al. 2004). For all stars, we started with an initial $T_{eff}$ and [Fe/H] of 3500 K and 0.00 dex respectively, and a corresponding $\log g$ of 0.65 obtained from the YY isochrones. Since the stars analyzed in this work are OC members, we assumed they belong to the thin disk population. Hence, we assumed a starting value of 0.00 dex for [O/Fe], which corresponds to solar metallicity for the thin disk. We then used PySME to fit the selected lines in the observed spectrum by varying the values of free parameters. The fits were performed employing $\chi^2$ minimization and as the output; we obtained new values of $T_{eff}$, [Fe/H], $v_{mic}$, $v_{mac}$, [C/Fe], and [N/Fe]. We then updated the value of log g from the YY isochrones corresponding to the $T_{eff}$ and [Fe/H] from the last PySME run. Similarly, the oxygen abundance was updated using the newly determined [Fe/H]. This cycle is repeated until there is a negligible difference between the values of all free parameters from the current PySME run and the previous PySME run. Further details of the method, as well as the comparison with a few benchmark M giant stars, are explained in Nandakumar et al. (2023a). Assuming an uncertainty of 0.15 dex in [O/Fe], they estimated typical uncertainties of ±100 K in $T_{eff}$, ±0.2 dex in log g, ±0.1 dex in [Fe/H], ±0.1 km s$^{-1}$ in $v_{mic}$, ±0.1 dex in [C/Fe], and ±0.1 dex in [N/Fe]. The final parameters for all our stars are given in Table 3.

| Stellar cluster | Star | Gaia DR3 ID | RA (deg) | Dec (deg) | G (mag) | S/N H-band | S/N K-band |
|---|---|---|---|---|---|---|---|
| NGC 7789 | N7789_1 | 1995061928762465536 | 359.263535573 | 56.766078036 | 9.75 | 359 | 327 |
| age=1.55Gyr | N7789_2 | 1995014409242207872 | 359.293345577 | 56.713695317 | 9.86 | 356 | 314 |
| $R_{gc}$ = 9.43 kpc | N7789_4 | 1995059592301326848 | 359.074303178 | 56.668800703 | 10.22 | 255 | 226 |
| NGC 7044 | N7044_1 | 1969807040026523008 | 318.321873119 | 42.484572587 | 11.79 | 481 | 404 |
| age=1.66Gyr | N7044_2 | 1969807276235623552 | 318.330484209 | 42.507969696 | 11.90 | 426 | 327 |
| $R_{gc}$ = 8.73 kpc | N7044_3 | 1969806073644788992 | 318.397775571 | 42.460809968 | 12.20 | 388 | 346 |
| | N7044_4 | 1969800576086654592 | 318.256942841 | 42.403479143 | 12.21 | 412 | 328 |
| NGC 6819 | N6819_a | 2076394728016615680 | 295.479787235 | 40.239351300 | 10.07 | 655 | 519 |
| age=2.24Gyr | N6819_b | 2076582950658667264 | 295.284268649 | 40.325501253 | 10.13 | 519 | 364 |
| $R_{gc}$ = 8.03 kpc | | | | | | | |
| Ruprecht 171 | Rup171_1 | 4103073693495483904 | 277.989813692 | -15.980947363 | 10.01 | 397 | 336 |
| age=2.75Gyr | Rup171_2 | 4102882309792631552 | 278.022115687 | -16.133756431 | 10.45 | 383 | 311 |
| $R_{gc}$ = 6.90 kpc | | | | | | | |
| Trumpler 5 | Trumpler5_1 | 3326783231129992704 | 99.253875467 | 9.496152447 | 10.72 | 382 | 359 |
| age=4.27Gyr | | | | | | | |
| $R_{gc}$ = 11.21 kpc | | | | | | | |
| King 11 | King11_1 | 2211216117949545216 | 356.988703461 | 68.595172524 | 11.66 | 372 | 355 |
| age = 4.47Gyr | King11_2 | 2211121972266402304 | 356.899458081 | 68.559189268 | 12.04 | 395 | 348 |
| $R_{gc}$ = 10.21 kpc | King11_3 | 2211220211058075776 | 356.910794815 | 68.656763219 | 12.24 | 195 | 183 |
| NGC 6791 | N6791_2 | 2051105616974709504 | 290.316961400 | 37.77952520 | 12.25 | 168 | 153 |
| age = 8.31Gyr | N6791_3 | 2051287002031070208 | 290.207176626 | 37.72853365 | 12.39 | 159 | 138 |
| $R_{gc}$ = 7.94 kpc | | | | | | | |

Table 1: Basic information of the program stars and spectra. The ages and $R_{gc}$ are from Cantat-Gaudin et al. (2020) except for NGC 6791 (Brogaard et al. 2021). The signal-to-noise ratios (S/N) were obtained from the fits header of each spectrum, using the value for order 48 as representative of the H-band and order 33 for the K-band.

| Species | $\lambda_{air}$ (Å) | log($gf$) | $E_{low}$ (eV) | Reference |
|---|---|---|---|---|
| HF | 22778.249 | -3.969 | 0.674 | Jönsson et al. (2014a) |
| HF | 23358.329 | -3.962 | 0.227 | Jönsson et al. (2014a) |
| Ce II | 16595.180 | -2.114 | 0.122 | Montelius et al. (2022) |
| Ce II | 17058.880 | -1.425 | 0.318 | Montelius et al. (2022) |

Table 2: Lines used to determine fluorine and cerium abundances.

### 3.2. Stellar abundances of F and Ce

Since the resolving power of the instrument varies as a function of wavelength, a second, slightly larger, $v_{mac}$ was determined from K-band Fe lines. In this way two "$v_{mac}$ determinations" have been carried out. The first, slightly smaller, from the stellar parameter determination from H-band lines, is used when determining Ce abundances from the H-band. The second, slightly larger adjusted using K-band lines, is used when determining F abundances using the K-band HF lines.

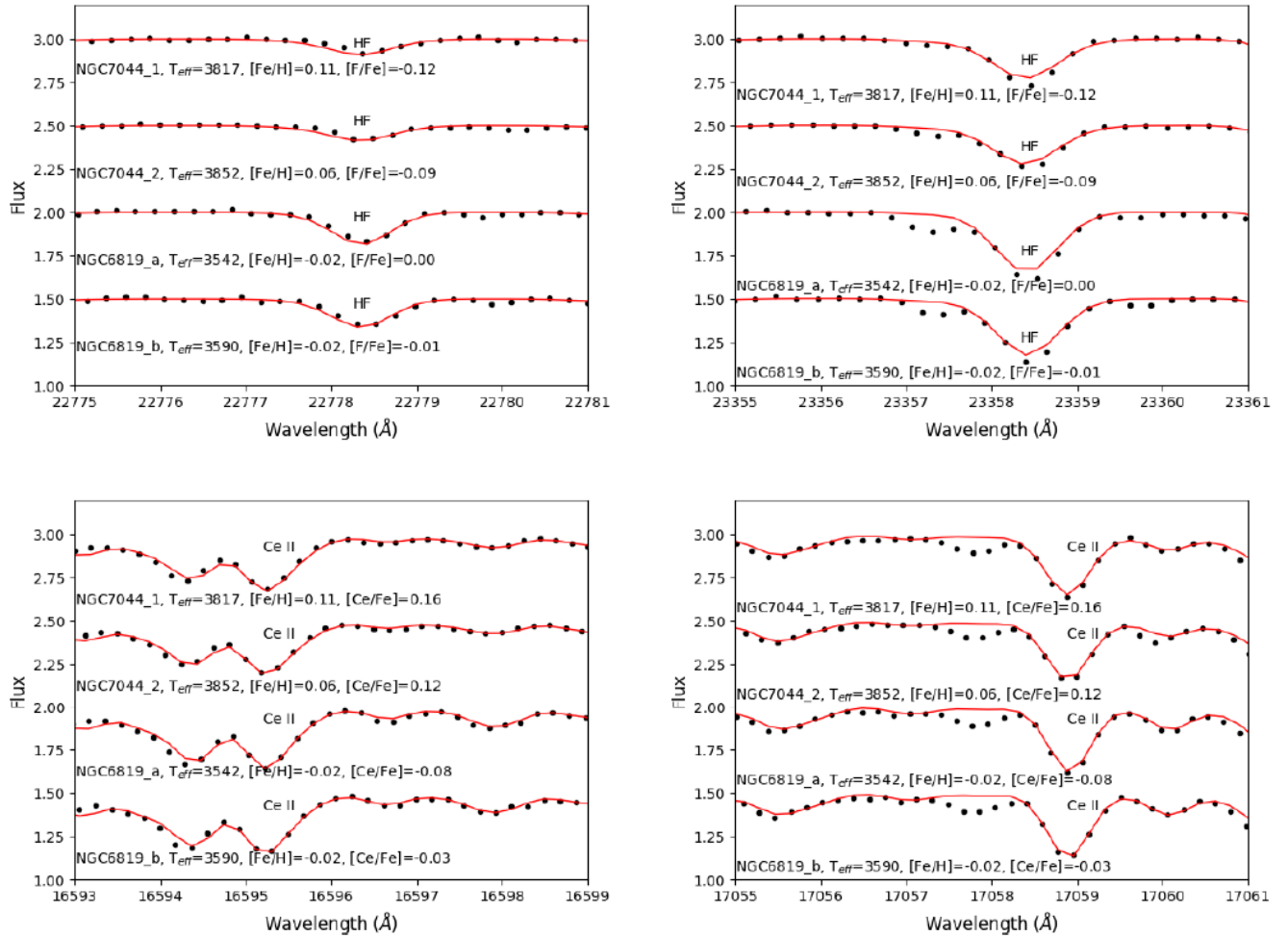

Fig. 1: Observed GIANO-B spectra of two stars each from NGC 7044 and NGC 6819. The HF lines shown in the spectra are 22778.249 Å and 23358.329 Å. The Ce II lines are 16595.180 Å and 17058.880 Å. The solid red line is the synthetic spectra, and the black dots are the observed spectra. The respective effective temperature, metallicity, and fluorine abundances are indicated. Only the HF lines and Ce II lines are fitted.

Uncertainties associated with all abundance measurements were assessed using a Monte Carlo technique. 100 sets of values of $T_{eff}$, log g, [Fe/H], and $v_{mic}$ were randomly produced and normally distributed around our determined parameters, based on the methodological uncertainties estimated in Nandakumar et al. (2023a). Then each spectrum was analyzed 100 times using these 100 different stellar parameters, resulting in 100 different abundance values of F and Ce.

As a result of this comprehensive analysis, we were able to provide a robust estimate of the uncertainties in the F and Ce abundances stemming from the uncertainties of the stellar parameters. The median absolute deviation of these individual star-by-star abundance uncertainties are listed as the abundance uncertainties for the individual stars in Table 3.

## 4. Results and discussion

The main results are listed in Tables 3 and 4, are shown in Fig. 2 and Fig. 3, and are discussed in detail below. We investigated the different relationships between [F/Fe], [Ce/Fe], [F/H], and [Ce/H] with age and $R_{gc}$. Our results are also compared with GCE models that implement different stellar sources of F and Ce in Fig. 4 and Fig. 5.

### 4.1. Comparison of the determined stellar parameters with literature values

The same stars in the OCs NGC 7044 and Ruprecht 171 were previously studied by Casali et al. (2020), who used two different methods to determine the stellar parameters. Specifically, they employed Fast Automatic MOOG Analysis (FAMA) (Magrini et al. 2013), which is an automated procedure based on an equivalent width analysis wrapping the MOOG code (Sneden 1973), and ROTFIT (Frasca et al. 2006), which performs a $\chi^2$ minimization of the entire spectrum. Both these tools use optical spectra, while our stellar parameter determination was made using H-band OH, CN, CO, and Fe lines.

For NGC 7044 we find an average metallicity [Fe/H]=0.08 ± 0.02 dex, whereas Casali et al. (2020) in-ferred $-0.37 \pm 0.14$ dex using FAMA and $-0.13 \pm 0.09$ dex by exploiting ROTFIT. For FAMA, they observe a clear trend for $T_{eff}$ and [Fe/H], which is not expected. This trend becomes particularly pronounced for stars with low temperatures and log g values, causing their metallicities to plummet to exceedingly low levels, thus rendering their reliability questionable. Conversely, the ROTFIT values align much more closely with our estimates.

In the case of Ruprecht 171, we find a global metallicity [Fe/H]=$0.03 \pm 0.01$ dex. In Casali et al. (2020) there are a total of seven stars. Using FAMA, they found the three coolest stars of Ruprecht 171 (Rup171_1, Rup171_2, and Rup171_3) to have a much lower metallicity than the other members of the same cluster. By considering only the warmer stars of Ruprech 171 in their sample, the metallicity is [Fe/H]=$-0.07 \pm 0.11$. This is still lower than, but more consistent with, our value, once observational uncertainties are considered. In contrast, when examining the ROTFIT estimates, including all seven stars spanning a range of temperatures from the lowest to the highest, the resulting average metallicity is found to be $0.03 \pm 0.1$, displaying remarkable agreement with our derived value.

For Trumpler 5, we have only one star and [Fe/H]=$-0.40 \pm 0.01$, but that is in good agreement with the mean cluster metallicity of Donati et al. (2015) ([Fe/H] = $-0.403 \pm 0.006$), Donor et al. (2020) ([Fe/H]=$-0.44 \pm 0.01$), and Lucertini et al. (2023) ([Fe/H]=$-0.49 \pm 0.14$). The two stars in NGC 6819 ([Fe/H]=$-0.02 \pm 0.01$) are in excellent agreement with Lee-Brown et al. (2015) who analyzed medium-resolution spectra using the Wisconsin-Indiana-Yale-NOIRLab (WIYN) for stars in NGC 6819 and found [Fe/H]=$-0.03 \pm 0.06$ dex. Our estimates agree well with the value of [Fe/H]=$0.03 \pm 0.04$ by Myers et al. (2022). But our value is slightly lower than Bragaglia et al. (2001), who estimated a value of [Fe/H]=$+0.09 \pm 0.03$.

Our determined metallicity of NGC 6791 averages at [Fe/H]=$+0.26 \pm 0.01$ dex, aligning well with the findings of Linden et al. (2017) with [Fe/H]= $+0.30 \pm 0.02$ and Villanova et al. (2018) with [Fe/H]= $+0.31 \pm 0.01$. However, it is lower than the one measured by Gratton et al. (2006) where they found [Fe/H]=$+0.47 \pm 0.12$. Remarkably, this cluster is not only the oldest in our collection, with an age of 8.31 Gyr, but also one of the most puzzling due to its extreme kinematic properties, such as tidally induced rotation (Kamann et al. 2019). Its old age, combined with a notably high metallicity, further contributes to its distinctiveness. The cluster's current position, R=7.94 kpc away from the Galactic Center, raises intriguing questions about the evolutionary pathways it undertook to attain such high metallicity levels. The unusual properties and possible origins of the cluster NGC 6791 have been thoroughly investigated in several studies. Carraro et al. (2006) observed a large eccentricity (0.59) for this OC and interpreted it as possibly being the core of a larger system undergoing intense tidal stripping. They also suggested that the cluster may have formed near the metal-rich bulge on the inner side of the Galaxy. Jílková et al. (2012) used simulations to investigate the cluster's orbit and possible migration processes. They integrated the orbit of NGC 6791 into a model of the Milky Way's gravitational potential, taking into account the effects of the Galaxy's halo, bulge, disk, bar, and spiral arms. Their results suggest that a model that included strong bar and spiral arm perturbations could explain the migration of NGC 6791 from an inner disk position (3-5 kpc from the Galactic Center) to its current location. Linden et al. (2017) hypothesized that NGC 6791 may be an intrinsic part of the thick disk, or may have originally belonged to the Galactic bulge. This hypothesis is supported by the high metal ([Fe/H] = 0.28 - 0.34) and high α (0.08 - 0.10) abundances found in five members analyzed using APOGEE DR13 data. Martinez-Medina et al. (2018) suggested that NGC 6791 formed in either the inner thin disk or the bulge, and later migrated to its present position. Similarly, Villanova et al. (2018) classified NGC 6791 as a member of the Galactic disk based on its spatial coordinates (z, RGC) = (1, 8) kpc. Their spectroscopic analysis, revealing [Fe/H] = $+0.313 \pm 0.005$ and [α/Fe] = $+0.06 \pm 0.05$ in giant stars, supports a scenario where the cluster originated in the Galactic bulge and underwent radial migration. Additionally, works using the *Gaia* results (Gaia Collaboration et al. 2018) confirmed the unusual properties of NGC 6791. For instance, Carrera et al. (2022) found an eccentricity of 0.35 for NGC 6791, suggesting that the cluster may have originated in the inner regions of the Galaxy, which is supported by its high metal content. Finally, Karataş et al. (2023) found migration and eccentricity values of $d_{mig}$ = 7.12 kpc and $ecc$ = 0.29 for NGC 6791 that support the radial migration scenarios. Nandakumar et al. (2024) discovered five metal-rich stars in the inner bulge with higher [F/Fe] abundances. This enhancement in F for old, metal-rich bulge stars is remarkably similar to NGC 6791, which is also old and metal-rich. This demonstrates that the trend of F with respect to [Fe/H] in the inner regions of the Galaxy differs from the trend observed in the solar neighborhood. Hence, our analysis corroborates NGC 6791's unique status, underscoring its deviation from the typical age-metallicity patterns observed in the Galactic disk as shown in Fig. 2. In their study, Nagarajan et al. (2023) derived a metallicity of [Fe/H]=$-0.02 \pm 0.05$ dex for NGC 7789, which is consistent with our findings of $0.00 \pm 0.01$ dex.

The good agreement with previous works underscores the reliability of our results. Finally, our work presents the first-ever reported metallicity for King 11 ([Fe/H] = $-0.25 \pm 0.01$), as it has not been previously studied with high-resolution spectroscopy. The reason for this is its high level of extinction, making it impractical to observe red clump stars, which are privileged targets of many OC studies (e.g., Bragaglia et al. 2001; Casamiquela et al. 2019). However, we intentionally targeted the brighter segment of the red giant branch since it is better suited for HF measurements.

### 4.2. F and Ce abundances

In Fig. 2 (upper panels) we plot [F/Fe] and [F/H] as a function of metallicity for our sample clusters together with the field stars from Nandakumar et al. (2023b). Some literature OCs (see the caption for details) are also used in the panels comparing [F/Fe] and [F/H] to [Fe/H], but also age and $R_{gc}$ in the second and third panels. The paper by Nandakumar et al. (2023b) presents a precise line-by-line abundance analysis of ten molecular HF lines in the Immersion GRating INfrared Spectrometer (IGRINS) spectra of 50 M giants, and examines the nature of the F trend as a function of metallicity. Fig. 2 shows that, except for NGC 6791, our values are in remarkable agreement with the abundance trend defined by field stars in Nandakumar et al. (2023b). In the investigations carried out by Ryde (2020) and Guerço et al.

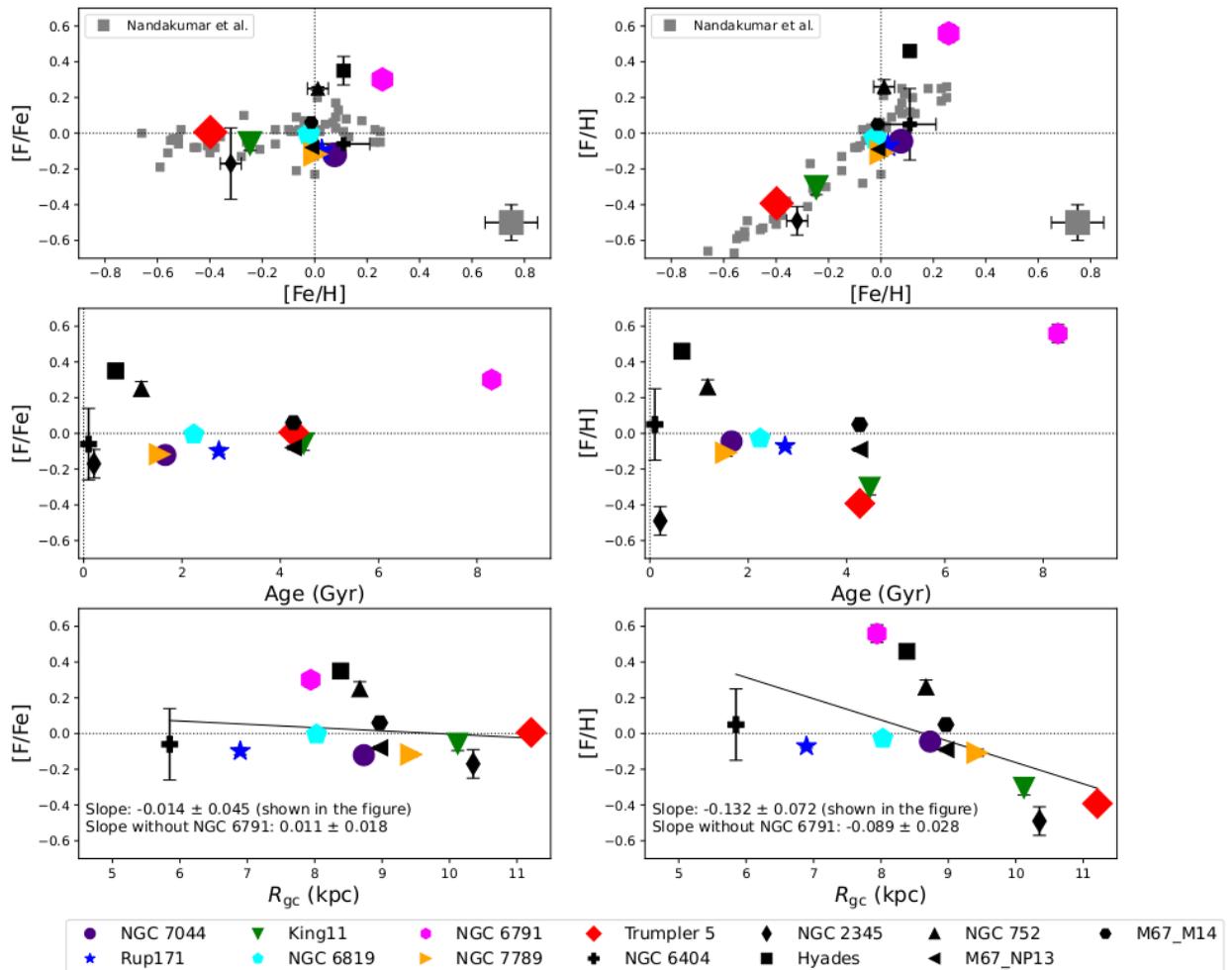

Fig. 2: Left: Relationship between [F/Fe] and [Fe/H], age, and $R_{gc}$. Right: Same but for [F/H]. Data from Nandakumar et al. (2023b) are depicted as gray dots. We also report values for five OCs, namely Hyades, NGC 752, and M67 from Nault & Pilachowski (2013), another estimate of M67 and another cluster, NGC 6404, from Maiorca et al. (2014), and NGC 2345 from Holanda et al. (2023). The slopes $-0.014 \pm 0.045$ and $-0.132 \pm 0.072$ (shown in the figure) were obtained by incorporating the calculations of NGC 6791 into the plot.

(2022), a secondary pattern in the behavior of F emerges at high, super-solar metallicities — a pattern absent in the findings from Nandakumar et al. (2023b). This discrepancy can be ascribed to the previous works using the relatively strong and possibly saturated/blended HF R9 line, which displays temperature-related trends and is found to show a high degree of uncertainty, particularly among cooler and metal-rich giant stars (for a comprehensive discussion, see Nandakumar et al. 2023b).

The second row of plots in Fig. 2 presents the variation of [F/Fe] and [F/H] as a function of the cluster age for our selected OCs, together with values reported in the literature for M67 and NGC 6404 (Maiorca et al. 2014), Hyades, NGC 752, M67 (Nault & Pilachowski 2013), and NGC 2345 (Holanda et al. 2023). This represents the entirety of fluorine measurements in OCs reported to date. The data for our clusters, together with NGC 2345, show a consistent flat trend between about 200 Myr and 5 Gyr, with the exception of NGC 6791, which shows an enhanced [F/Fe] ratio, a peculiarity given its age. Although NGC 6404 appears to have a slight excess of [F/Fe], the uncertainties involved are too large to confirm this conclusively. In particular, the Hyades and NGC 752 stand out as anomalous in this distribution. This may be due to the fact that the lines analyzed in these relatively warm stars are weak and difficult to analyze.

Finally, we considered [F/Fe] and [F/H] ratios as a function of the galactocentric distances in the bottom row of panels in Fig. 2. We see a flat trend for [F/Fe] versus $R_{gc}$, while there is a declining trend for [F/H] with $R_{gc}$, as would be expected according to the Galactic metallicity gradient. This pattern holds for all OCs, except for NGC 6791. When calculating the slope for [F/H] versus $R_{gc}$, we find a value of $-0.132 \pm 0.072$ dex/kpc and $-0.089 \pm 0.028$ dex/kpc with and without NGC 6791, respectively.

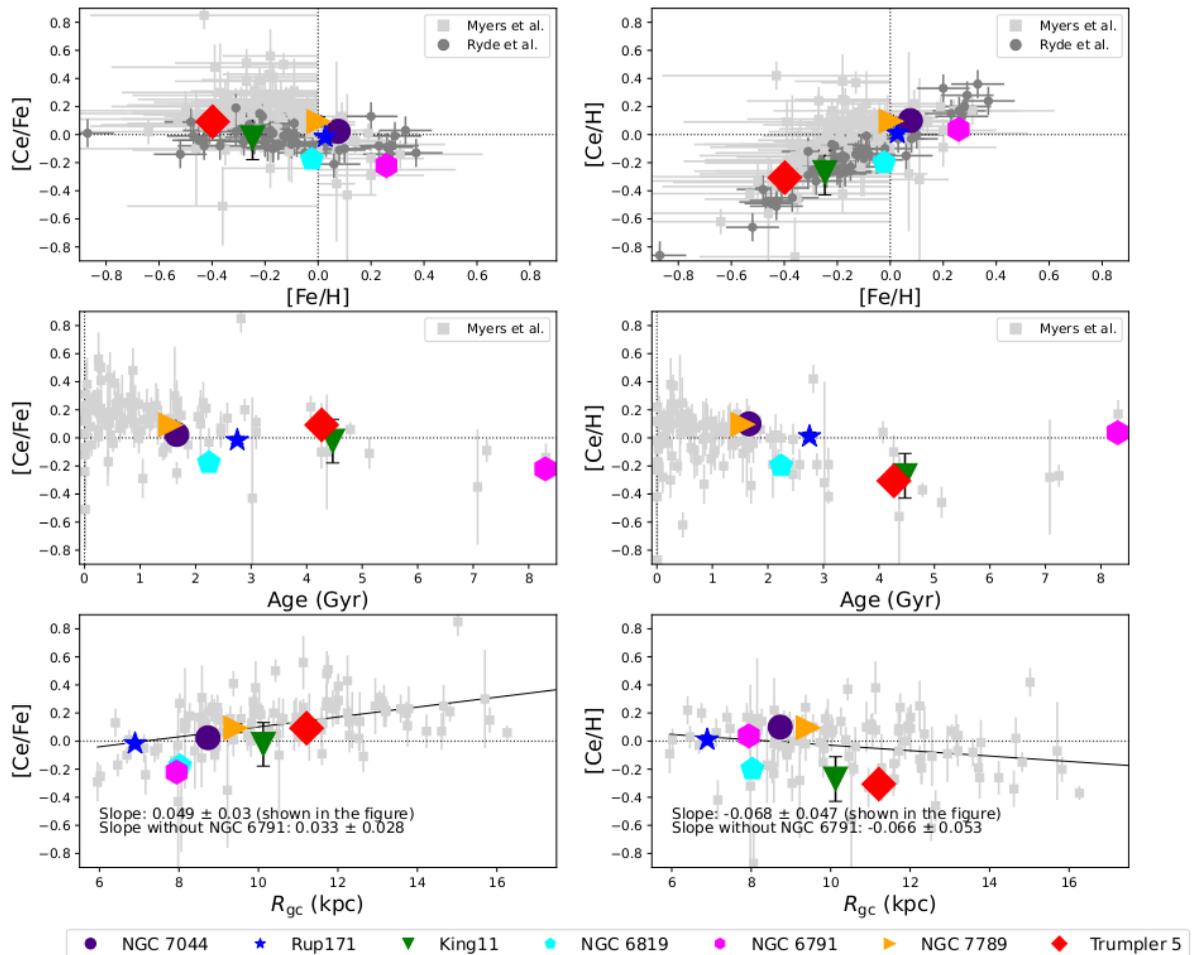

Fig. 3: Relationship between [Ce/Fe] and [Fe/H], age, and $R_{gc}$ (left) and [Ce/H] as a function of [Fe/H], age, and $R_{gc}$ (right). Data for field stars taken from Ryde et al. (2020) are shown as gray dots. The clusters taken from Myers et al. (2022) are shown as light gray squares. The slopes $0.049 \pm 0.03$ and $-0.068 \pm 0.047$ (shown in the figure) were obtained by incorporating the calculations of NGC 6791 into the plot.

Fig. 3 in the same way illustrates the behavior of cerium with metallicity, age, and galactocentric distances. Myers et al. (2022) presented chemical abundances and radial velocities for OC stars from the APOGEE survey (DR17; Abdurro'uf et al. 2022). Their analysis encompassed a total of 150 OCs, from which we selectively included clusters equipped with available data on ages, distances, and [Ce/Fe] values in this plot, resulting in a subset of 104 OCs for direct comparison with our results. We have four clusters (NGC 6791, NGC 6819, Trumpler 5, and NGC 7789) in common with Myers et al. (2022). Additionally, we considered the data from 51 giant field stars presented by Ryde et al. (2020) in our comparative analysis. Our obtained values exhibit good general agreement with those reported by Ryde et al. (2020) and Myers et al. (2022), as depicted in the first row of subplots in Fig. 3. Also, a cluster-by-cluster comparison with the common clusters shows that our results agree, within the uncertainties, with those of Myers et al. (2022); their determined [Ce/Fe] values are $-0.14 \pm 0.1$, $-0.03 \pm 0.1$, $0.01 \pm 0.1$, and $0.15 \pm 0.1$ for NGC 6791, NGC 6819, NGC 7789, and Trumpler 5, respectively.

Upon examining the variations in [Ce/Fe] and [Ce/H] relative to the age of the clusters in the second row of plots in Fig. 3, we observe a decreasing trend, indicating that younger clusters tend to exhibit higher cerium abundance compared to their older counterparts. The trend of [Ce/Fe] with age corresponds consistently with the results of previous studies on s-process elements in OCs, demonstrating an increasing trend at younger ages (D'Orazi & Randich 2009; Maiorca et al. 2011; Yong et al. 2012; Mishenina et al. 2015; D'Orazi et al. 2022; Magrini et al. 2023). The value for NGC 6791 aligns well with the findings of Myers et al. (2022) but stands out conspicuously in the context of the [Ce/H] trend with age.

In the bottom row of plots in Fig. 3, we present the behavior of [Ce/Fe] and [Ce/H] as a function of galactocentric distances ($R_{gc}$). We calculated the slope for these plots both including and excluding the NGC 6791 cluster. For [Ce/Fe] versus $R_{gc}$, the calculated slope is $0.049 \pm 0.031$ dex/kpc, demonstrating excellent agreement with the findings of Myers et al. (2022); when NGC 6791 is excluded from the analysis, this values becomes $0.033 \pm 0.028$ dex/kpc. The values are $-0.068 \pm 0.047$ with NGC 6791 and $-0.066 \pm 0.053$ without NGC 6791 for [Ce/H] versus $R_{gc}$. The rising trend of [Ce/Fe] as a function of galactocentric distance also confirms previous results on OCs in *Gaia*-ESO by, for example, Viscasillas Vázquez et al. (2022) and Magrini et al. (2023).

### 4.3. Chemical evolution models

A comprehensive study conducted by Ryde et al. (2020) revealed intriguing trends in the relationship between [F/Fe] and [Fe/H]. They observed a relatively flat trend for [F/Fe] versus [Fe/H] when considering subsolar metallicities, but this trend became positive for [Fe/H] surpassing the solar value. This behavior, confirmed by Guerço et al. (2022), was difficult to explain by conventional mechanisms for fluorine production. Spitoni et al. (2018) suggested the possible involvement of novae and W-R channels, as they could be factors contributing to this intriguing phenomenon.

However, as discussed in Nandakumar et al. (2023b), a large part of this observed secondary behavior of [F/Fe] with [Fe/H] at solar and super-solar metallicities likely can be attributed to the previously described problems with the R9 HF line. Our results are possibly in line with those of Nandakumar et al. (2023b) (although NGC 6791 reveals a high [F/Fe] ratio) and do not confirm the upward trend of Ryde et al. (2020) at [Fe/H] > 0.15 dex. On the other hand, we confirm the flat trend of [F/Fe] versus [Fe/H] also seen in Ryde (2020); Guerço et al. (2022) in the metallicity range -0.6 < [Fe/H] < 0.1 dex.

In Figs. 4 to 6, our data are plotted together with two sets of models of the chemical evolution of the Galaxy with various stellar sources included, in the left and right-hand panels, respectively. In Fig. 4, [F/Fe], [F/H] are plotted against [Fe/H], while in Fig. 5, [F/Fe], [F/H] are plotted against age. NGC 6791 continues to be a curious outlier when compared to these models of the chemical evolution of the local disk. The ratios of [Ce/Fe] and [Ce/H] plotted against [Fe/H] and age, respectively, are shown in 6.

The adopted chemical evolution models assume that the thick- and thin-disk components form out of two distinct episodes of gas accretion (two-infall model). Spitoni et al. (2019) updated the classical two-infall model proposed by Chiappini et al. (1997), considering constraints on stellar ages from asteroseismology (Silva Aguirre et al. 2018). They fixed the delay time for the second infall to a much higher value than assumed in the original formulation. In particular, in this work, we used the following model parameters: infall timescales for the thick and thin-disk components $\tau_1 = 1$ Gyr and $\tau_2 = 7$ Gyr, respectively. The delay time for the second infall is 3.25 Gyr. We assume the Kroupa et al. (1993) IMF and the star formation rate (SFR) are expressed as the Kennicutt (1998) law:

$$\psi = \nu \sigma_g^k, \quad (1)$$

where $\sigma_g$ is the gas surface density and $k = 1.5$ is the exponent. The quantity $\nu$ is the star formation efficiency (SFE), which was set to 2 Gyr$^{-1}$ and 1 Gyr$^{-1}$ in the thick- and thin-disk phase, respectively. According to McKee et al. (2015), the total surface mass density in the solar neighborhood is 47 M$_\odot$ pc$^{-2}$. We consider values of 7 and 40 M$_\odot$ pc$^{-2}$ for the thick- and thin-disk components, respectively (Spitoni et al. 2020). Finally, no Galactic winds were considered. This choice was motivated by previous studies by Spitoni et al. (2008, 2009) on the Galactic fountains (processes initiated by the explosions of SNe II in OB associations). They found that metals fall back onto the disk to the proximity of the galactocentric region from which they were expelled, thereby having a negligible impact on the overall chemical evolution of the Galactic disk.

In the left panels of Fig. 4 and Fig. 5, we consider for the fluorine production the same nucleosynthesis sources as model F7 of Spitoni et al. (2018): AGB stars (Karakas 2010), SNe II (Kobayashi et al. 2006), Type Ia SNe (Iwamoto et al. 1999) and W-R stars (Meynet & Maeder 2002). To account for the observational uncertainties associated with the determination of Type Ia SN rates, we explored different values of the free parameter of the model that describes the fraction of the stellar mass ending up in binary systems with the right properties[1] to give rise to SNe Ia (cost=0.025, 0.035, and 0.045 cases in the figures). This fraction is assumed to be constant in space and time in the Galaxy.

In Fig. 4, the dilution effect caused by the second infall of primordial gas when the thin-disk phase begins can be seen in the [F/Fe] versus [Fe/H] abundance ratios. The late accretion of pristine gas has the effect of decreasing the metallicity of the stellar populations born immediately after the infall event, producing the nearly horizontal evolutionary track at roughly constant [F/Fe] ranging approximately from [Fe/H] ~ -0.3 dex to [Fe/H] ~ 0.25 dex. In this scenario, the metallicity does not show a monotonic increase throughout the evolutionary time. The end of the thick disk phase can be pinpointed as the point at which [Fe/H] starts decreasing in the [F/Fe]–[Fe/H] plane. This transition occurs after ~3.25 Gyr of Galactic evolution. We note that the predicted [F/Fe] versus [Fe/H] and [F/H] versus [Fe/H] abundance ratios (left-handed panels of Fig. 4) reproduce the observed ones satisfactorily. Furthermore, the [F/Fe] temporal evolution traced by the models (left upper panel of Fig. 5) also agrees with the data. However, the [F/H] versus age relation is not properly reproduced. Finally, when considering only F injected into the ISM by AGB stars (red lines), it becomes evident that AGB stars predominantly contribute to fluorine production (adopting Spitoni et al. (2018) nucleosynthesis prescriptions).

The right-hand panels of the plots show the results that we obtained by implementing other, more recent stellar nucleosynthesis prescriptions in the chemical evolution code, to assess the theoretical uncertainties associated with the use of different yield sets. The right-hand panels of Fig. 4 and Fig. 5 display, respectively, the predicted [F/Fe] and [F/H] ratios as functions of [Fe/H] and age when the following combinations of yields are used: For massive stars (including W-R stars), we considered three yield sets, i.e., set $R$ of Limongi & Chieffi (2018) for nonrotating ($v_{\rm rot}$ = 0 km s$^{-1}$) and fast-rotating stars ($v_{\rm rot}$ = 150 and 300 km s$^{-1}$); for low- and intermediate-mass stars (LIMSs), we con-sidered either the yields from the FUll-Network Repository of Updated Isotopic Tables and Yields (FRUITY) database (Cristallo et al. 2009, 2011, 2015) or the yields computed by the Monash team (Lugaro et al. 2012; Fishlock et al. 2014; Karakas & Lugaro 2016; Karakas et al. 2018). As regards the FRUITY yields, we chose the yields computed with standard assumptions about the extent of the $^{13}$C pocket. As regards the MONASH yields, we also chose models with standard $^{13}$C pockets, following Karakas & Lugaro (2016). As a general rule, standard sizes of the pockets are as follows: $2 \times 10^{-3}$ M$_\odot$ for stellar masses $M \leq 3$ M$_\odot$, $1 \times 10^{-3}$ M$_\odot$ for stellar masses in the range 3–4 M$_\odot$, $1 \times 10^{-4}$ M$_\odot$ for stellar masses in the range 4–5 M$_\odot$, no $^{13}$C pocket for $M \geq 5$ M$_\odot$. These choices reflect the fact that the size of the He-intershell becomes smaller in mass with increasing stellar mass. The FRUITY yields are computed for model stars with initial masses in the range 1–6 M$_\odot$. The MONASH yields are computed for stars with initial masses from 0.9–1 M$_\odot$ to 6–8 M$_\odot$, with the exact mass range depending on the initial metallicity of the set of stellar models. We performed simple linear interpolations and extrapolations in the cases where stellar yields were not tabulated.

---

[1] They include the total mass, secondary-to-primary stellar mass ratio, separation, and mass accretion rate following Matteucci (1986) and Greggio & Renzini (1983)).

In the right-handed panels of Fig. 4 and Fig. 5 we show two models that include F production from AGB stars only (dashed lines; see the figure legend and caption). It appears that MONASH F yields almost suffice to explain the observational data points while using the FRUITY yields requires the addition of considerable F production in the hydrostatic burning layers of massive fast rotators. The yields of fast-rotating massive stars by Limongi & Chieffi (2018), however, appear to produce too much F. It has been shown that to reproduce the trends of the abundance ratios of several elements in the Galaxy, the bulk of massive stars must rotate faster (more slowly) at low (high) metallicities (Prantzos et al. 2018; Romano et al. 2019). Therefore, we considered a model (cyan continuous line in Fig. 4) in which 80% of low-metallicity massive stars are fast rotators, while the remaining do not rotate at all. On the other hand, at solar and super-solar metallicities these fractions reverse, with a smooth transition between the two regimes. This model can better explain the observations. We also notice that it is likely that the average rotational velocities of the stars are lower than about 150 km s$^{-1}$. Prantzos et al. (2018) consider an initial distribution of rotational velocities depending on the initial metallicity of the stars and interpolate the existent grids of yields. However, the behavior of the yields as a function of rotational velocity does not appear to be monotonic, and the grid in $v_{\rm rot}$ is very sparse, so we prefer to avoid this kind of interpolation.

Cerium yields are highly dependent on assumptions about the size of the $^{13}$C pocket in AGB stars. We followed the suggestion of Molero et al. (2023), halving the yields to obtain a reasonable fit to the data when using the FRUITY set of yields. However, we also show a model (green dashed lines in Fig. 6) in which no correction is applied to the yields. From Fig., reffig:h it is seen that a better fit to the observations for Ce can be obtained by halving the Ce yields provided in the FRUITY database (FRUITY/2; i.e., with corrections), while also accounting for Ce production from massive stars with either $v_{\rm rot} = 150$ km s$^{-1}$ or 0 km s$^{-1}$ (see the pink and blue lines in Fig. 6); see also Molero et al. (2023). The AGB nucleosynthesis calculations of the Monash team appear to overestimate the yields of Ce as well. Hopefully, our results can be used to improve the models of stellar evolution and nucleosynthesis.

## 5. Summary and concluding remarks

In this paper, we have presented a comprehensive analysis of the stellar parameters, metallicities, and fluorine and cerium abundances for a total of 17 stars distributed across seven OCs: NGC 7044, NGC 7789, Ruprecht 171, Trumpler 5, King 11, NGC 6819, and NGC 6791. Our primary focus has been to unravel the intricate trends governing the abundances of cerium and fluorine as a function of clusters metallicity, distance from the Galactic Center, and age. Particularly noteworthy is our pioneering work in deriving the radial metallicity gradient for fluorine using OCs. We find a flat trend with metallicity that provides invaluable insights into the different channels of F production and its evolution over time. These multiple proposed channels in the formation and evolution of F encompass a wide range of potential sources, from low-mass AGB stars to W-R stars, SNe II, and novae. As found in previous works, the old, metal-rich cluster NGC 6791 remains an outlier in our F analysis, including when compared to chemical evolution models tailored for the solar neighborhood. The inclusion of more super-solar-metallicity clusters in the sample may be the key to a better understanding of the nature and origin of this anomalous cluster.

A comparison with chemical evolution models shows that the results from the FRUITY database, which includes LIMSs (Cristallo et al. 2009, 2011, 2015), is under-producing F compared to our sample stars, and therefore another source is needed. However, the rotating massive stars of Limongi & Chieffi (2018) overproduce F. On the other hand, the MONASH yields from LIMSs calculated by Lugaro et al. (2012); Fishlock et al. (2014); Karakas & Lugaro (2016); Karakas et al. (2018) and the F7 model of Spitoni et al. (2019) give a reasonable fit for our sample stars when [F/H] is compared with metallicity, but not so much when compared with age. We obtained a satisfactory fit to the data in all diagnostic planes by assuming that F is produced by AGB stars (adopting MONASH yields) and by massive stars, with the fraction of fast rotators varying as a function of metallicity (on average, there must be more fast rotators as the metallicity decreases; see also Prantzos et al. 2018; Romano et al. 2019).

The interplay between s-process elements and F is of great importance. In particular, AGB stars and massive stars can produce both s-process elements and F. Chemical evolution models that implement yields from both AGB stars and rotating massive stars overpredict the [Ce/Fe] and [Ce/H] ratios. A reasonable agreement between model predictions and observations can be obtained by lowering the Ce yields from AGB stars (e.g., by lowering the FRUITY yields by a factor of 2; see Molero et al. 2023) and by considering a combination of massive star yields with, for example, $v_{\rm rot} = 150$ km s$^{-1}$ and 0 km s$^{-1}$, as proposed here.

Although a mix of massive stars with different initial $v_{\rm rot}$ values along with AGB yields offers a viable solution, to disentangle the situation it would be beneficial to trace individually a variety of s-process elements individually. Further work into F abundances across different cluster ages, with a particular emphasis on younger and older clusters, is also essential. While AGB stars are currently considered

| Star | T$_{eff}$ (K) | log$g$ (dex) | [Fe/H] (dex) | v$_{mic}$ (km s$^{-1}$) | v$_{mac}$ (km s$^{-1}$) | [C/Fe] (dex) | [N/Fe] (dex) | [O/Fe] (dex) | [F/Fe] (dex) | [F/H] (dex) | [Ce/Fe] (dex) | [Ce/H] (dex) |
|---|---|---|---|---|---|---|---|---|---|---|---|---|
| N7789_1 | 3811 | 1.20 | 0.00 | 1.76 | 6.30 | -0.18 | 0.26 | -0.04 | -0.13±0.11 | -0.13±0.15 | 0.09±0.11 | 0.09±0.15 |
| N7789_2 | 3802 | 1.20 | 0.02 | 1.68 | 5.96 | -0.18 | 0.21 | -0.05 | -0.10±0.11 | -0.09±0.15 | 0.05±0.07 | 0.07±0.12 |
| N7789_4 | 3930 | 1.43 | 0.00 | 1.55 | 6.28 | -0.18 | 0.25 | -0.04 | -0.12±0.14 | -0.13±0.17 | 0.13±0.09 | 0.13±0.16 |
| N7044_1 | 3817 | 1.30 | 0.11 | 1.75 | 6.78 | -0.19 | 0.26 | -0.08 | -0.12±0.08 | -0.02±0.13 | 0.04±0.08 | 0.15±0.13 |
| N7044_2 | 3852 | 1.32 | 0.06 | 1.72 | 6.93 | -0.17 | 0.27 | -0.06 | -0.09±0.11 | -0.03±0.15 | 0.00±0.07 | 0.06±0.12 |
| N7044_3 | 3890 | 1.39 | 0.06 | 1.66 | 7.02 | -0.20 | 0.26 | -0.06 | -0.12±0.13 | -0.05±0.16 | 0.02±0.10 | 0.08±0.14 |
| N7044_4 | 3877 | 1.39 | 0.08 | 1.67 | 6.61 | -0.19 | 0.24 | -0.07 | -0.16±0.13 | -0.08±0.16 | 0.03±0.07 | 0.11±0.12 |
| N6819_a | 3542 | 0.71 | -0.02 | 1.83 | 5.85 | -0.14 | 0.15 | -0.03 | 0.00±0.10 | -0.03±0.14 | -0.20±0.08 | -0.22±0.13 |
| N6819_b | 3590 | 0.79 | -0.02 | 1.82 | 6.16 | -0.14 | 0.11 | -0.03 | -0.01±0.08 | -0.03±0.13 | -0.15±0.08 | -0.17±0.13 |
| Rup171_1 | 3881 | 1.36 | 0.02 | 1.60 | 6.77 | -0.16 | 0.23 | -0.05 | -0.08±0.12 | -0.06±0.16 | -0.04±0.10 | -0.02±0.14 |
| Rup171_2 | 4012 | 1.62 | 0.03 | 1.51 | 6.72 | -0.17 | 0.19 | -0.06 | -0.12±0.12 | -0.08±0.16 | 0.01±0.08 | 0.03±0.13 |
| Trumpler5_1 | 3582 | 0.58 | -0.40 | 1.99 | 6.98 | -0.11 | 0.19 | 0.10 | 0.01±0.09 | -0.39±0.13 | 0.09±0.08 | -0.31±0.13 |
| King11_1 | 3356 | 0.28 | -0.25 | 1.90 | 10.25 | -0.17 | 0.21 | 0.05 | -0.04±0.07 | -0.29±0.12 | -0.18±0.09 | -0.43±0.13 |
| King11_2 | 3711 | 0.86 | -0.25 | 1.85 | 6.39 | -0.11 | 0.13 | 0.05 | -0.11±0.10 | -0.36±0.14 | -0.01±0.07 | -0.26±0.12 |
| King11_3 | 4083 | 1.56 | -0.24 | 1.57 | 6.78 | -0.13 | 0.21 | 0.05 | -0.02±0.14 | -0.26±0.17 | 0.11±0.08 | -0.13±0.13 |
| N6791_2 | 3493 | 0.83 | 0.25 | 1.81 | 6.50 | 0.05 | 0.10 | 0.06 | 0.26±0.10 | 0.51±0.14 | -0.24±0.08 | 0.01±0.13 |
| N6791_3 | 3490 | 0.87 | 0.27 | 1.76 | 6.39 | 0.02 | 0.14 | 0.06 | 0.34±0.11 | 0.61±0.15 | -0.21±0.08 | 0.06±0.13 |

Table 3: Parameters and abundances of the program stars. Abundance ratios are scaled to the solar values of A(C)$_\odot$ = 8.46, A(N)$_\odot$ = 7.83, A(O)$_\odot$ = 8.69, A(F)$_\odot$ = 4.40, A(Fe)$_\odot$ = 7.46, and A(Ce)$_\odot$ = 1.58 (Asplund et al. 2021). Typical uncertainties of the derived stellar parameters are ±100 K in T$_{eff}$, ±0.2 dex in log $g$, ±0.1 dex in [Fe/H], ±0.1 km s$^{-1}$ in v$_{mic}$, ±0.1 dex in [C/Fe], and ±0.1 dex in [N/Fe].

| Stellar cluster | Age (Gyr) | R$_{GC}$ (kpc) | [Fe/H] (dex) | [F/Fe] (dex) | [F/H] (dex) | [Ce/Fe] (dex) | [Ce/H] (dex) |
|---|---|---|---|---|---|---|---|
| NGC 7789 | 1.55 | 9.43 | 0.00±0.01 | -0.11±0.01 | -0.11±0.02 | 0.10±0.03 | 0.10±0.03 |
| NGC 7044 | 1.66 | 8.73 | 0.08±0.02 | -0.12±0.02 | -0.04±0.02 | 0.02±0.03 | 0.10±0.03 |
| NGC 6819 | 2.24 | 8.03 | -0.02±0.01 | 0.00±0.01 | -0.03±0.01 | -0.18±0.02 | -0.20±0.02 |
| Ruprecht 171 | 2.75 | 6.90 | 0.03±0.01 | -0.10±0.02 | -0.07±0.01 | -0.02±0.02 | 0.01±0.03 |
| Trumpler 5 | 4.27 | 11.21 | -0.40±0.01 | 0.01±0.01 | -0.39±0.01 | 0.09±0.01 | -0.31±0.01 |
| King 11 | 4.47 | 10.12 | -0.25±0.01 | -0.06±0.04 | -0.30±0.04 | -0.02±0.12 | -0.27±0.12 |
| NGC 6791 | 8.31 | 7.94 | 0.26±0.01 | 0.30±0.04 | 0.56±0.05 | -0.22±0.02 | 0.04±0.03 |

Table 4: Averaged abundance values for the seven stellar clusters. The uncertainties quoted above are calculated from the root mean square deviation from the mean of the values determined for individual stars within that cluster, except for Trumpler 5, where we only have one star in our sample. For this cluster, the listed uncertainties instead are representative estimates based on the mean uncertainties of the other clusters.

the primary source of F, the observed anomalies strongly suggest the involvement of other sources.

*Acknowledgements.* We are grateful to the anonymous reviewer for their careful and constructive feedback, which has significantly enhanced the quality of this paper. VD acknowledges the financial contribution from PRIN-MUR 2022YP5ACE. Financial support from Mini-Grant INAF 2022 (High-Resolution Observations of Open Clusters) is acknowledged. G.N. acknowledges the support from the Wenner-Gren Foundations (UPD2020-0191 and UPD2022-0059) and the Royal Physiographic Society in Lund through the Stiftelsen Walter Gyllenbergs fond. This work has made extensive use of the Simbad, Vizier, and NASA ADS databases.

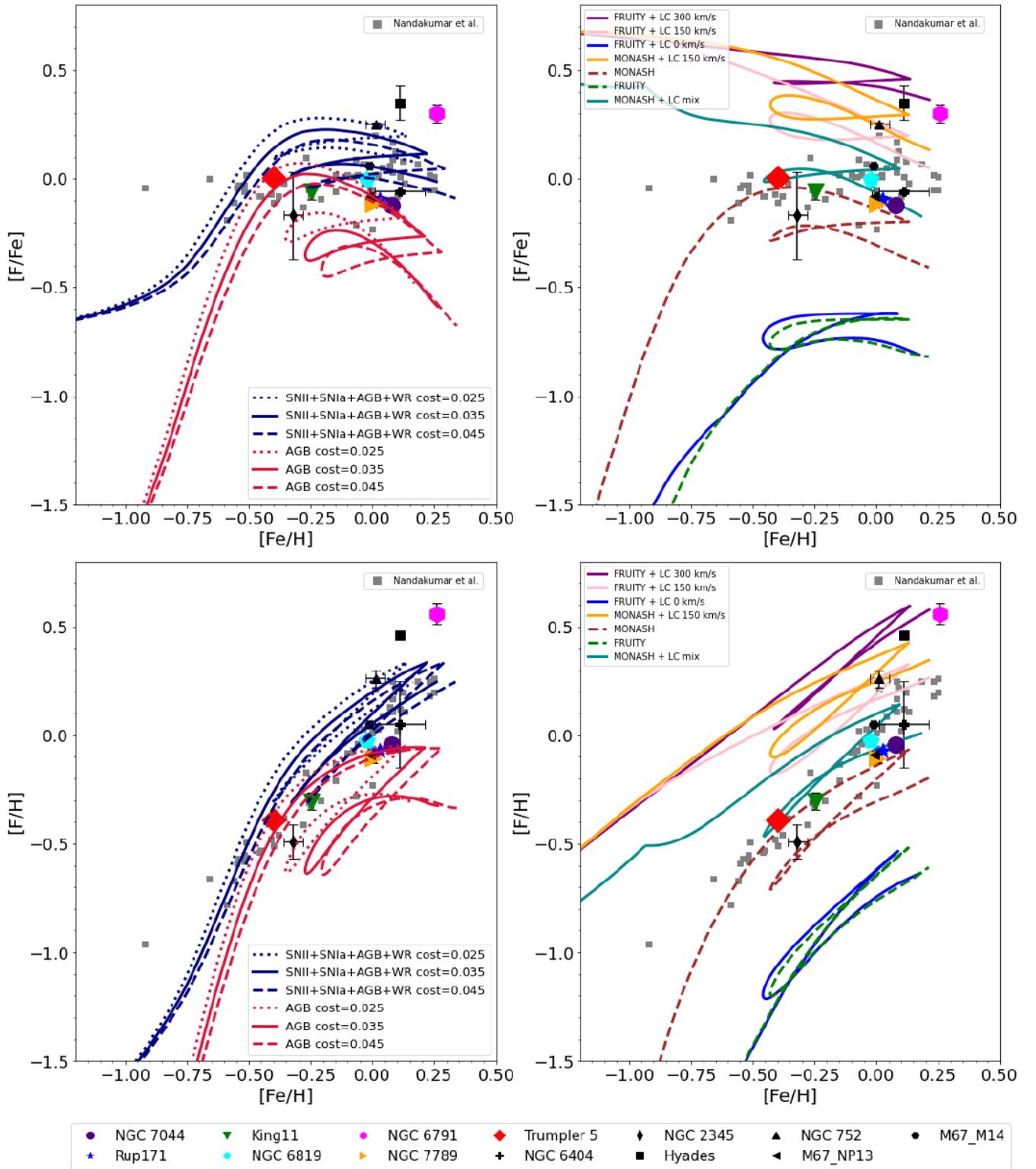

Fig. 4: [F/Fe] and [F/H] versus [Fe/H]. The colored dots represent our sample clusters, and the lines are based on theoretical models. In the left panel, the solid lines are based on the two-infall Spitoni et al. (2019) chemical evolution models, which assume the same fluorine nucleosynthesis prescription as model F7 of Spitoni et al. (2018) (see Sect 4.3. for further details). The right panel shows the models that implement the production of F from both LIMSs and massive stars (solid lines) and the models that implement the production of F from AGB stars only (dashed lines).

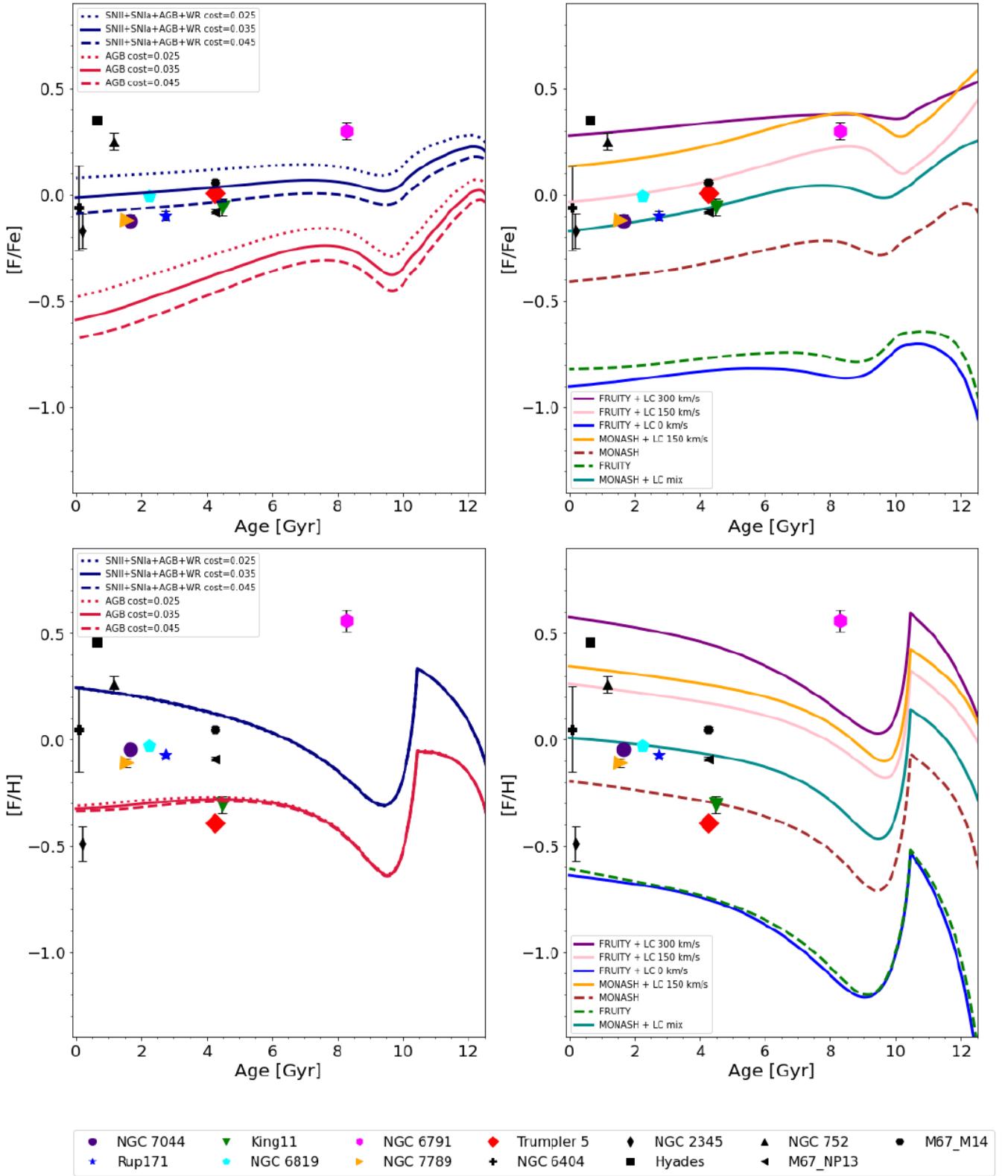

Fig. 5: [F/Fe] and [F/H] versus age. The data and the models are the same as in Fig. 4.

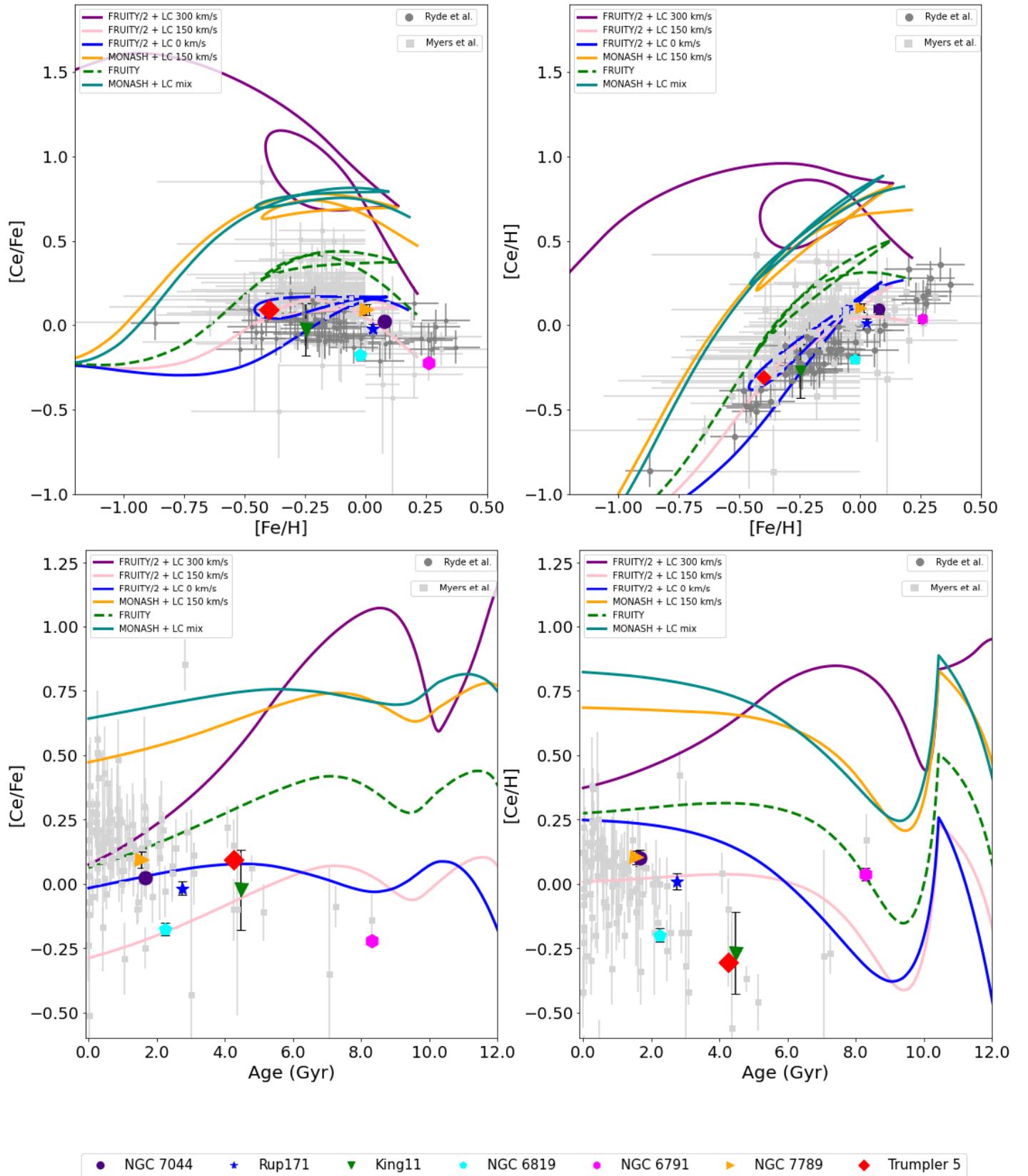

Fig. 6: [Ce/Fe] and [Ce/H] versus [Fe/H] and age for our sample stars. The lines shown in the plots are the same Romano et al (in prep.) models as in Fig. 4 and Fig. 5, but for Ce.